# Orbital memory from individual Fe atoms on black phosphorus

*Brian Kiraly[†], Elze J. Knol[†], Alexander N. Rudenko[†], Mikhail I. Katsnelson[†], Alexander A. Khajetoorians[†*]*

[†]Institute for Molecules and Materials, Radboud University, Nijmegen 6525AJ, The Netherlands

[*] Correspondence to: a.khajetoorians@science.ru.nl

Bistable valency in individual atoms presents a new approach toward single-atom memory, as well as a building block to create tunable and stochastic multi-well energy landscapes. Yet, this concept of orbital memory has thus far only been observed for cobalt atoms on the surface of black phosphorus, which are switched using tip-induced ionization. Here, we show that individual iron atoms on the surface of black phosphorus exhibit orbital memory using a combination of scanning tunneling microscopy and spectroscopy with *ab initio* calculations based on density functional theory. Unlike cobalt, the iron orbital memory can be switched in its non-ionized ground state. Based on calculations, we confirm that each iron valency has a distinct magnetic moment that is characterized by a distinguishable charge distribution due to the different orbital population. By studying the stochastic switching of the valency with varying tunneling conditions, we propose that the switching mechanism is based on a two-electron tunneling process.

**Introduction**

Atomic arrays of coupled orbital memory have been used to demonstrate tunable multiplicity, namely multi-well energy landscapes which can be controllably shaped.[1, 2] Using scanning tunneling microscopy/spectroscopy (STM/STS), the observation of a tunable multiplicity was evidenced by stochastic multi-state noise with vastly different time scales generated from Co atoms on black phosphorus (BP)[2]. This observation was used to mimic a Boltzmann machine at the atomic level, based on the long-range nature of the interactions between individual atoms.[3] Coupled arrays of magnetic atoms on surfaces that exhibit orbital memory are in stark contrast to arrays of magnetic atoms on other surfaces coupled *via* exchange mechanisms.[4-7] In these examples, the local and strong nature of the exchange interactions leads to strongly favorable ordered states, often observable in two-state telegraph noise. Understanding the role of spin in orbital memory, and the presence of other time scales is fundamentally important toward understanding glassy dynamics[1] and may provide a new route toward tuning exchange interactions.[8]

In order to explore tunable multiplicity, it is imperative to understand the physical mechanisms responsible for orbital memory and the nature of its switching mechanism. Orbital memory in individual atoms is based on the concept of bistable valency, where two distinct values of the atomic magnetic moment are stable and reversible. Orbital memory and the switching between two distinct magnetic moments is reminiscent of high/low spin transitions seen in individual molecules [9], or so-called spin crossover molecules, however scaled down to an individual atom. While a multi-stable valency was first discussed for individual Co atoms on graphene/graphite,[10-14] it has thus far only been experimentally observed for individual Co atoms on BP.[2] For Co on BP, the interplay between the population of the various hybrid orbitals and dielectric screening leads to a state-dependent spatially distributed charge density and distinct relaxations, which can be directly imaged with STM/STS. This mechanism is distinct than the aforementioned spin crossover molecules, where transitions are also strongly linked to conformational changes of the molecule. Complementary, it was shown that individual Co atoms can be ionized due to tip-induced band banding (TIBB), similar to other impurities on semiconductor surfaces, and used to switch between the orbital memory states.[2, 15] Sufficiently large TIBB leads to stochastic switching between the two ionized orbital states, where the occupation lifetimes can be strongly tuned by the selected tunneling conditions. The influence of electrostatic fields was further illustrated in recent experiments where it was shown that the electrostatic field generated by a local Cu dopant strongly modified the stochastic switching behavior of an individual Co atom, distinctive from the local TIBB. The sensitivity of Co to local electrostatic fields and the absence of current-induced switching in the unionized state makes it extremely challenging to explore the nature of the switching mechanism as well as the role of spin in orbital memory, motivating the search for orbital memory in other materials.

In this work, we demonstrate orbital memory from individual Fe atoms on BP using a combination of scanning tunneling microscopy/spectroscopy (STM/STS) and *ab initio* calculations based on density functional theory (DFT). Using constant-current imaging, we observe two stable configurations of an individual Fe atom residing in a hollow site, denoted $Fe_{low}$ and $Fe_{high}$, which are electrically reversible and possess distinguishable charge densities. Based on our calculations, we attribute these two different charge densities to a charge redistribution between the 4*s* and 3*d* orbitals of the Fe atom, where each configuration has a distinct spin moment ($m_{low} \sim 2\mu_B$ and $m_{high} \sim 3.5\mu_B$). By comparing the experimental data with *ab* *initio* calculations of the electronic structure, we link spectroscopic features to hybrid atomic orbitals. We also observe stochastic switching between $Fe_{low}$ and $Fe_{high}$, where the relative favorability of the two states and their residence lifetimes can be tuned by the applied tunneling conditions. However, unlike previous observations for Co on BP, this stochastic switching is observed in the ground state, i.e. without ionization, and is seen at both bias polarities. By specifically studying the current-dependent response of the residence lifetimes, we propose that the switching mechanism is governed by a two-electron process.

**Experimental observation of orbital memory for individual Fe atoms**

After deposition, the signature of individual Fe atoms is manifested by the appearance of an isotropic species of adsorbates with an apparent height of $252 \pm 3$ pm (Figure 5a). Based on atomic imaging of the BP lattice, we identified the binding site of the deposited Fe atoms as a top site. Tunneling into individual atoms above voltages of $V_s = 700$ mV causes the apparent height and shape of the atoms to change irreversibly (Figure 5b,c), which we confirm is a binding site shift from a top to hollow site. When imaging an individual hollow-site Fe atom at various tunneling conditions, the apparent height and spatial distribution can be switched reversible between two distinct and stable configurations, $Fe_{low}$ and $Fe_{high}$ (Fig. 1a,b). We detail the tunneling depending switching further below. Binding site analysis (Figure 6) reveals that both observed configurations reside in the same hollow site of the BP lattice. The species can further be distinguished *via* their apparent height (measured with $V_s = -400$ mV), with $\Delta z_{low} = 92 \pm 2$ pm and $\Delta z_{high} = 59 \pm 2$ pm. We therefore, attribute $Fe_{low}$ and $Fe_{high}$ to two distinct orbital configurations.

**Theoretical/experimental characterization of the electronic structure of Fe atoms**

In order to identify the electronic structure of $Fe_{low}$ and $Fe_{high}$, we performed DFT and DFT+U calculations[16-24]. The calculations confirm the existence of two distinct valencies, contingent upon the Hubbard-U parameter used in the calculations, reminiscent of orbital memory.[2] The results for low-U calculations (from U = 0 eV to U = 1.5 eV) show a larger *d*-orbital occupation and smaller spin moment compared to the high-U calculations (U > 2 eV) (table 1, see also Figure 7). Integrating the total density of states (DOS) over an energy window near the valence band edge, we calculated the spatial distribution of the charge density for both $Fe_{low}$ and $Fe_{high}$ (Fig. 1c,d). We observed qualitative agreement between the STM constant-current image shown in Fig. 1a (Fig. 1b) and the calculated charge density distribution in Fig. 1c (Fig. 1d), enabling us to confirm that each configuration can be attributed to a distinct and stable valency of the Fe atom. We also calculated the relaxed vertical positions of the Fe atoms ($d_{low}$ = 109 pm and $d_{high}$ = 142 pm), which are pictorially depicted in Fig. 1e,f. Consistent with the picture of screening based orbital repopulation,[2] the lower spin state resides closer to the BP surface than the higher spin state.

To further characterize the electronic properties of the Fe atom, d$I$/d$V$ spectra of $Fe_{low}$ were taken (Fig. 2(a)). Unfortunately, the valency switches unidirectional from $Fe_{high}$ to $Fe_{low}$ near $|V_s|$ = 500 mV, strongly favoring $Fe_{low}$, precluding the possibility of collecting a d$I$/d$V$ spectrum (at $|V_s|$ > 500 mV) for $Fe_{high}$. The spectrum in Fig. 2(a) shows two resonances at $V_s$ = -700 mV and $V_s$ = 580 mV. The resonances have large linewidths (FWHM = 160 mV) and the peak positions are insensitive to the setpoint conditions of the spectra. This allows us to rule out the influence of tip-induced band bending (Figure 8), which is distinctly different compared to Co on BP. Examining the total DOS for $Fe_{low}$ (Figure 2b), we see several hybrid *d*-orbitals below the Fermi energy. As the DFT calculations include only one monolayer of BP, we cannot make a direct comparison of the theoretical and experimental peak energies; however, we can note that the peak calculated at approximately -0.5 eV is composed primarily of out-of-plane orbitals and should thus be most apparent in STM/STS measurements. Furthermore, calculations show that this peak is also spin polarized (Figure 9). Several peaks also appear in the total DOS above the $E_F$ (Figure 2b). In the energy range from 0.3 eV to 0.8 eV, both Fe $s$-orbitals and $d$-orbitals hybridize with the $p$-orbitals of the BP, which most likely accounts for the origin of the peak seen in STS at $V_s$ = 580 mV. Finally, the calculated DOS for Fe$_{high}$ indicates that the Fe 3$d$ states are generally shifted away from the Fermi energy, a common consequence of a higher U parameter.

**Stochastic switching of individual Fe atoms**

Above an applied DC threshold voltage ($|V_s| > 380$ mV), we observed stochastic switching as evidenced by telegraph noise in the measured tunneling current ($I_t$) (Fig. 3a,b). We define the state-dependent residence times as $\tau_{low}$ and $\tau_{high}$, the mean residence lifetime as $\tau_m = (\tau_{low} + \tau_{high})/2$, and the lifetime asymmetry as $A = (\tau_{low} - \tau_{high})/(\tau_{low} + \tau_{high})$. In order to extract $\tau_{low}$, $\tau_{high}$, and $\tau_m$, we measured telegraph noise in both constant height (Figure 3) and constant current (Figure 4) on the highest point (the center) of the Fe atom. We measured a minimum of 1000 switches at a given set of tunneling conditions. We then utilize the procedure described in ref. [25] to extract $\tau_{low}$, $\tau_{high}$, and $\tau_m$. We observed, as shown in Fig. 3c, that increasing $|V_s|$ leads to a monotonous decrease of $\tau_m$. As we discuss below, a similar trend is observed with increasing $I_t$. In Fig. 3e,f, we clearly see that the favorability between Fe$_{low}$ and Fe$_{high}$ depends on both the magnitude and polarity of $V_s$. Given $V_s < 0$, it can be seen that the magnitude of $V_s$ strongly modifies the asymmetry. At $V_s$ < -460 mV, Fe$_{low}$ is more favorable ($A > 0$), with its favorability increasing with increasing magnitude, while the situation is reversed for $V_s$ > -460 mV ($A < 0$) and the overall asymmetry is close to 0. The qualitative behavior of the switching is dramatically modified at positive polarities, namely for $V_s > 0$, as seen in Fig. 3b. In this bias regime, the residence lifetime of Fe$_{high}$ is strongly suppressed, with $A \approx 1$ for all measured values of $V_s$ (Fig. 3f). As with the opposite bias polarity, $\tau_m$ (gray curve, Fig. 3d) decreases with increasing $V_s$ (and $I_t$).

When examining the switching efficiency as a function of tip position, the Fe only showed appreciable switching rates with the tip positioned within roughly 500 pm from the atomic center (Figure 10). This is in stark contrast to the Co atom, which switched with the tip as far as 2 nm away from the center of the atom.[2] We attribute the difference to the role of tip-induced band bending. For the case of Fe, the switching occurs in the neutral state, whereas Co is first ionized via tip-induced band bending and switching is observed near and above voltages where this occurs.

To gain further insight into the switching mechanism between Fe$_{low}$ and Fe$_{high}$, we studied the dependence of the mean switching frequency ($\nu_m = 1/\tau_m$) on the magnitude of the tunneling current ($I_t$) (Figure 4a). Telegraph noise was acquired with the tip parked on the center of the Fe atom at a fixed voltage with the feedback on to maintain constant current. As seen in Fig. 4a, $\nu_m$ increases monotonously with increasing $I_t$. When plotted with a logarithmic scale on both axes, $\nu_m$ shows a power-law dependence with respect to the tunneling current: $\nu_m = I_t^N$. Fits to the experimental data (Fig. 4b), reveal that $N \approx 2$ for the measured range of $I_t$.

Switching processes with a power-law dependence on the tunneling current have been observed in studies of individual atoms and molecules in STM-based experiments, e.g. in atom manipulation, molecule dissociation, molecule rotation[26] and hydrogen tautomerization[27, 28]. In these experiments, the associated switching rate has a power-law dependence, $\nu = I_t^N$. Here, $N$ is related to the number of events that are needed to overcome the energy barrier associated with the switching process. It was shown theoretically, that the required energy can be provided via inelastic tunneling processes. For Fe on BP, the $N \approx 2$ power-law dependence suggests that there is an intermediate state that mediates switching of the valency, i.e. overcoming the energy barrier. There are multiple processes that can potentially link an intermediate state to facilitate switching. Quasiparticle excitations involving the lattice which are activated via inelastic tunneling, for example phonons or polarons, can induce vertical motion of Fe atom. An induced motion of the Fe atom may lead to a non-zero probability to switch to the opposing orbital state, due to the distinct relaxation height of each valency state. Such motion would otherwise be frozen out at our measurement temperature. However, phonon-driven excitations via inelastic tunneling should be symmetric with respect to bias polarity, and alone could not explain the polarity-dependent switching behavior. The observed switching may also be induced by local charging of the Fe atom. While we do not observe evidence of charging, the lifetime associated with charge occupation of the Fe atom may be longer than the flux associated with the tunneling current. Such a mechanism would imply a charge-mediated switching mechanism and would potentially require tunneling electrons with sufficient energy to overcome to charge occupational energy associated with single charge occupation. Based on the current experimental data, we cannot rule out such potential mechanisms. However, it is most likely that the ground-state energy barrier between the two valency states is larger than the energy needed to activate such processes; otherwise, we would expect the switching process to be directly excited by inelastic tunneling yielding a $N \approx 1$ power law.

**Conclusion**

We have shown that Fe adsorbed in the hollow site of BP exhibits orbital memory, namely two distinct and stable valency configurations. These two memory bearing states can be read out from the distinct charge densities each configuration exhibits, and can be switched electrically. The valency configurations of Fe lead to two distinct magnetic moments, and the electronic structure indicates signatures of the hybrid atomic orbitals. Moreover, Fe orbital memory exhibits two distinct regimes: a) a regime where the given valency is frozen, and b) a regime where the orbital states can be driven stochastically with tunneling current, as evidenced by telegraph noise. This finding illustrates that the concept of orbital memory is not unique to Co on BP. The stochastic switching of Fe can be driven at both voltage polarities, which is contrary to Co on BP. Likewise, the stochastic switching of the Fe atom can be switched without ionizing the atom, as there was no clear experimental evidence of significant tip-induced band bending on the atomic states. Based on this observation, we probed the switching frequency with tunneling current which revealed a clear power law suggesting that the switching mechanism is based on a two-electron process. Finally, the spatially-dependent, bipolar stochastic switching in the Fe atom strongly contrasts the Co, introducing the possibility to add complexity into coupled ensembles of orbital memory.

**Acknowledgements**

This project has received funding from the European Research Council (ERC) under the European Union's Horizon 2020 research and innovation program (SPINAPSE: Grant agreement No. 818399). E. J. K. and A. A. K. acknowledge support from the NWO-VIDI project 'Manipulating the interplay between superconductivity and chiral magnetism at the single-atom level' with project no. 680-47-534. B. K. acknowledges the NWO-VENI project 'Controlling magnetism of single atoms on black phosphorus' with project no. 016.Veni.192.168. For the theoretical part of this work, A. N. R. and M. I. K. received funding from the European Research Council via Synergy Grant 854843 – FASTCORR.

**Appendix A: Materials and methods**

The STM and STS measurements were performed with a commercial Omicron STM in ultrahigh vacuum ($p < 1\times10^{-10}$ mbar) and at temperature of $T = 4.4$ K. The bias $V_s$ was applied to the sample. All measurements were performed with electrochemically etched W tips, prepared *in situ* by electron bombardment and field emission. The tips were dipped and characterized on a clean Au(111) surface, before approaching a BP sample. The BP crystals were purchased from HQ graphene. To obtain an atomically clean surface, the crystals were cleaved with scotch tape in UHV directly before their transfer to the STM for *in situ* characterization. Single Fe atoms were evaporated directly into the microscope with $T_{sample} < 5$ K during the dosing procedure. The STM images in this work were all acquired using constant-current feedback. STS measurements were performed using a lock-in technique to directly measure d$I$/d$V$. For the stochastic switching measurements that were acquired with the tip at constant height, the tip height was stabilized on the bare BP at $I_t = 20$ pA, $V_s = -400$ mV, before opening the feedback loop. In constant height telegraph noise measurements, the tip height was set by stabilizing on the BP surface ($V_{set} = -400$ mV and $I_{set} = 20$ pA) before measuring at the specified $V_s$ in the center of the Fe atom.

DFT calculations were carried out using the projected augmented-wave method (PAW)[16] as implemented in the Vienna *ab initio* simulation package (VASP).[17, 18] Exchange and correlation effects were taken into account within the spin-polarized generalized gradient approximation (GGA-PBE).[19] Additional Hubbard-U correction was applied to the 3d shell of Fe within the GGA + U method[23] in order to capture the effect of the distance-dependent Coulomb screening. An energy cutoff of 300 eV for the plane-wave basis and the convergence threshold of $10^{-6}$ eV were used, which is sufficient to obtain numerical accuracy. Pseudopotentials were taken to include 3s and 3p valence electrons for P atom, as well as 3d, 3p and 4s valence electrons for Fe atoms. The BP surface was modeled in the slab geometry by a single BP layer with dimensions $(3a \times 4b) \approx (13.1 \times 13.3)$ Å with atomic positions fixed to the experimental parameters of bulk BP.[29] Vertical separation between the layers was set to 20 Å. The Brillouin zone was sampled by a uniform distribution of *k*-points on an $(8 \times 8)$ mesh. The position of the Fe atom was relaxed considering top and hollow surface sites as starting points. We checked that the inclusion of two additional BP layers in the slab does not significantly affect the results. The primary difference between the single-layer and three-layer slabs is the reduction of a gap between the valence and conduction BP states. The behavior of Fe atom remains virtually unchanged including the adsorption distances, charge density distribution, and magnetic moments.

The projection of the electronic bands on specific atomic states was performed using the formalism of Wannier functions [21] implemented in the wannier90 package.[22] The spatial charge density distributions shown in Fig. 1c-d and Supplementary Figure 7 were calculated by performing band decomposition of the total DFT charge densities and averaging them over the energy interval of ~0.3 eV in the valence band edge, similar to ref.[2] The resulting charge distribution reflects the surface charge densities typical to the low-energy hole states in BP with Fe adatoms and, therefore, can be associated with the experimental STM images shown in Fig. 1a-b. This procedure is closely related to the Tersoff-Hamann scheme.[24] Possible mismatch between the simulated and experimental STM images can be related to the effect of the STM tip, which is not explicitly considered in this approach.

**Appendix B: Fe/BP sample characterization**

Figure 5a shows a typical sample after deposition of Fe atoms. The characteristic phosphorus vacancies are identified as dumbbell shaped protrustions with an apparent height that depends on their depth.[30] Three types of Fe atoms are identified. The tallest species is dominant after deposition and has an apparent height of $252 \pm 3$ pm. Analogous to Co atoms on BP,[2] this is assumed to be Fe adsorbed in a top site, $Fe_T$. Using a bias pulse of approximately $|V_s| > 700$ mV, top site Fe atoms can irreversibly be pushed into a hollow site (Figure 5b,c).

We can identify two Fe species in a hollow site, with different apparent heights of $92 \pm 2$ pm and $59 \pm 2$ pm. We can assign these species to different valencies $Fe_{low}$ ($92 \pm 2$ pm) and $Fe_{high}$ ($59 \pm 2$ pm), where the labels low and high refer to the magnetic moment. To confirm that the binding site of the two different hollow site species is identical, binding site analysis was performed using atomic resolution constant-current STM images (Figure 6). After determining the positions of the phosphorus atoms in the topmost monolayer (green gridlines in Figure 6 show every other atom in the zig-zag rows), the geometric center of the Fe atom was determined using the contour plots in the lowest panels of Figure 6. As seen in Figure 6a,b, the Fe atom in both valencies resides in precisely the same hollow binding site. Furthermore, when comparing another atom (Figure 6c,d), it is clear than again both valencies also reside in the same binding site, which is the mirror symmetric site to the atom in Figure 6a,b. The binding site symmetry is reflected in the constant-current STM images shown in the top panels of Figure 6, and was also observed for Co and Cu atoms on BP.[2, 15]

**Appendix C: Charge densities for Fe as a function of Hubbard-U**

To demonstrate that the two orbital configurations occur for a range of Hubbard-U parameter values, we plot the total energy of the $Fe_{low}$ and $Fe_{high}$ state as a function of the Hubbard-U parameter in Figure 7. Here, it is seen that for $U < 1.8$ eV the $Fe_{low}$ state is energetically favorable, and for $U > 1.8$ eV, the $Fe_{high}$ state is favorable. This difference is also reflected in the calculated charge density, seen in the lower panels of Figure 7. It is seen that the shape of the charge density is qualitatively similar for $U > 1.8$ eV, as well as for $U < 1.8$ eV.

**Appendix D: Insensitivity to tip-induced band bending**

To understand the origin of the observed peaks in the d$I$/d$V$ spectra on the $Fe_{low}$ atom, we studied the influence of the tip height on those same d$I$/d$V$ spectra. As seen in Figure 8, bringing the tip closer to the atom during the d$I$/d$V$ measurement (using the initial setpoint current) does not strongly influence the position of the observed peaks in the d$I$/d$V$ spectra. The insensitivity of the peaks to the tip height indicates a minimal influence from tip-induced band bending, confirming that they do not originate from ionization events. Hence, we can attribute the peaks to local hybridized orbitals.

**Appendix E: Electronic structure of Fe$_{low}$ and Fe$_{high}$**

Calculations of the spin-resolved band structure and density of states (DOS) for the Fe on BP system are shown in Figure 9. The Fe$_{low}$ atom has multiple primarily *d*-hybrid orbitals below the Fermi energy (Figure 9a). These hybrid states correspond to the peak observed at $V_s$ = -700 mV in the d$I$/d$V$ spectra taken on the low-spin state (Figure 2 / Figure 8). Furthermore, as seen in the spin-up vs. spin-down calculations, this state is also highly spin-polarized. Conversely, when examining the characteristics of Fe$_{high}$ atom, we find that the Fe *d*-orbitals are pushed quite far from the Fermi energy.

**Appendix F: Position-dependent switching**

Figure 10 shows the sensitivity of the telegraph noise to the precise tip position. The state-resolved residence times and mean residence times as a function of position along the [010] and [100] directions are plotted. In both directions, the residence times decrease rapidly when approaching the center of the Fe atom, as visible in the upper panels. As the tip moves away from the center of the Fe atom, the switching probability decreases significantly. This is in stark contrast to Co / BP, where the switching persists to up to approximately 3.5 nm from the atom in the [100] direction[2].

**Appendix G: Hydrogenation of Fe atoms**

In our experiments hydrogenated Fe atoms were only observed in the top binding site (see Fig. 11). In one given study, where we have co-deposited both Co and Fe, our observations of Fe indicate that Fe is less sensitive to long-term hydrogenation compared to Co (all top site).[2] In an analysis of nearly 750 atoms, we found that nearly 43% of the Co was hydrogenated, while only 8% of the Fe was hydrogenated.

**References**

1. Kolmus, A.; Katsnelson, M. I.; Khajetoorians, A. A.; Kappen, H. J., Atom-by-atom construction of attractors in a tunable finite size spin array. *New Journal of Physics* **2020,** *22* (2), 023038.
2. Kiraly, B.; Rudenko, A. N.; van Weerdenburg, W. M. J.; Wegner, D.; Katsnelson, M. I.; Khajetoorians, A. A., An orbitally derived single-atom magnetic memory. *Nature Communications* **2018,** *9* (1), 3904.
3. Kiraly, B.; Knol, E. J.; van Weerdenburg, W. M. J.; Kappen, H. J.; Khajetoorians, A. A., An atomic Boltzmann machine capable of self-adaption. *Nature Nanotechnology* **2021,** *16* (4), 414-420.
4. Hirjibehedin Cyrus, F.; Lutz Christopher, P.; Heinrich Andreas, J., Spin Coupling in Engineered Atomic Structures. *Science* **2006,** *312* (5776), 1021-1024.
5. Khajetoorians Alexander, A.; Wiebe, J.; Chilian, B.; Wiesendanger, R., Realizing All-Spin–Based Logic Operations Atom by Atom. *Science* **2011,** *332* (6033), 1062-1064.
6. Khajetoorians, A. A.; Wiebe, J.; Chilian, B.; Lounis, S.; Blügel, S.; Wiesendanger, R., Atom-by-atom engineering and magnetometry of tailored nanomagnets. *Nature Physics* **2012,** *8* (6), 497-503.
7. Toskovic, R.; van den Berg, R.; Spinelli, A.; Eliens, I. S.; van den Toorn, B.; Bryant, B.; Caux, J. S.; Otte, A. F., Atomic spin-chain realization of a model for quantum criticality. *Nature Physics* **2016,** *12* (7), 656-660.
8. Badrtdinov, D. I.; Rudenko, A. N.; Katsnelson, M. I.; Mazurenko, V. V., Control of magnetic interactions between surface adatoms via orbital repopulation. *2D Materials* **2020,** *7* (4), 045007.
9. Kahn, O., Spin-crossover molecular materials. *Current Opinion in Solid State and Materials Science* **1996,** *1* (4), 547-554.
10. Wehling, T. O.; Lichtenstein, A. I.; Katsnelson, M. I., Transition-metal adatoms on graphene: Influence of local Coulomb interactions on chemical bonding and magnetic moments. *Physical Review B* **2011,** *84* (23), 235110.
11. Rudenko, A. N.; Keil, F. J.; Katsnelson, M. I.; Lichtenstein, A. I., Adsorption of cobalt on graphene: Electron correlation effects from a quantum chemical perspective. *Physical Review B* **2012,** *86* (7), 075422.

12. Virgus, Y.; Purwanto, W.; Krakauer, H.; Zhang, S., Ab initio many-body study of cobalt adatoms adsorbed on graphene. *Physical Review B* **2012,** *86* (24), 241406.
13. Virgus, Y.; Purwanto, W.; Krakauer, H.; Zhang, S., Stability, Energetics, and Magnetic States of Cobalt Adatoms on Graphene. *Physical Review Letters* **2014,** *113* (17), 175502.
14. Sessi, V.; Stepanow, S.; Rudenko, A. N.; Krotzky, S.; Kern, K.; Hiebel, F.; Mallet, P.; Veuillen, J. Y.; Šipr, O.; Honolka, J.; Brookes, N. B., Single 3d transition metal atoms on multi-layer graphene systems: electronic configurations, bonding mechanisms and role of the substrate. *New Journal of Physics* **2014,** *16* (6), 062001.
15. Knol, E. J.; Kiraly, B.; Rudenko, A. N.; van Weerdenburg, W. M. J.; Katsnelson, M. I.; Khajetoorians, A. A., Gating orbital memory with an atomic donor. *arXiv e-prints* **2021**, arXiv:2107.07143.
16. Blochl, P. E., Projector Augmented-Wave Method. *Physical Review B* **1994,** *50* (24), 17953-17979.
17. Kresse, G.; Furthmuller, J., Efficient iterative schemes for ab initio total-energy calculations using a plane-wave basis set. *Physical Review B* **1996,** *54* (16), 11169-11186.
18. Kresse, G.; Joubert, D., From ultrasoft pseudopotentials to the projector augmented-wave method. *Physical Review B* **1999,** *59* (3), 1758-1775.
19. Perdew, J. P.; Burke, K.; Ernzerhof, M., Generalized gradient approximation made simple. *Physical Review Letters* **1996,** *77* (18), 3865-3868.
20. Brown, A.; Rundqvist, S., Refinement of the crystal structure of black phosphorus. *Acta Crystallographica* **1965,** *19* (4), 684--685.
21. Marzari, N.; Mostofi, A. A.; Yates, J. R.; Souza, I.; Vanderbilt, D., Maximally localized Wannier functions: Theory and applications. *Reviews of Modern Physics* **2012,** *84* (4), 1419-1475.
22. Mostofi, A. A.; Yates, J. R.; Lee, Y.-S.; Souza, I.; Vanderbilt, D.; Marzari, N., wannier90: A tool for obtaining maximally-localised Wannier functions. *Computer Physics Communications* **2008,** *178* (9), 685-699.
23. Dudarev, S. L.; Botton, G. A.; Savrasov, S. Y.; Humphreys, C. J.; Sutton, A. P., Electron-energy-loss spectra and the structural stability of nickel oxide: An LSDA+U study. *Physical Review B* **1998,** *57* (3), 1505-1509.
24. Tersoff, J.; Hamann, D. R., Theory of the scanning tunneling microscope. *Physical Review B* **1985,** *31* (2), 805-813.
25. Khajetoorians, A. A.; Baxevanis, B.; Hübner, C.; Schlenk, T.; Krause, S.; Wehling, T. O.; Lounis, S.; Lichtenstein, A.; Pfannkuche, D.; Wiebe, J.; Wiesendanger, R., Current-Driven Spin Dynamics of Artificially Constructed Quantum Magnets. *Science* **2013,** *339* (6115), 55.
26. Stipe, B. C.; Rezaei, M. A.; Ho, W., Inducing and Viewing the Rotational Motion of a Single Molecule. *Science* **1998,** *279* (5358), 1907.
27. Liljeroth, P.; Repp, J.; Meyer, G., Current-Induced Hydrogen Tautomerization and Conductance Switching of Naphthalocyanine Molecules. *Science* **2007,** *317* (5842), 1203-1206.
28. Pan, S.; Fu, Q.; Huang, T.; Zhao, A.; Wang, B.; Luo, Y.; Yang, J.; Hou, J., Design and control of electron transport properties of single molecules. *Proceedings of the National Academy of Sciences* **2009,** *106* (36), 15259.
29. Brown, A.; Rundqvist, S., Refinement of the crystal structure of black phosphorus. *Acta Crystallographica* **1965,** *19* (4), 684-685.

30. Kiraly, B.; Hauptmann, N.; Rudenko, A. N.; Katsnelson, M. I.; Khajetoorians, A. A., Probing Single Vacancies in Black Phosphorus at the Atomic Level. *Nano Letters* **2017,** *17* (6), 3607-3612.

Figures

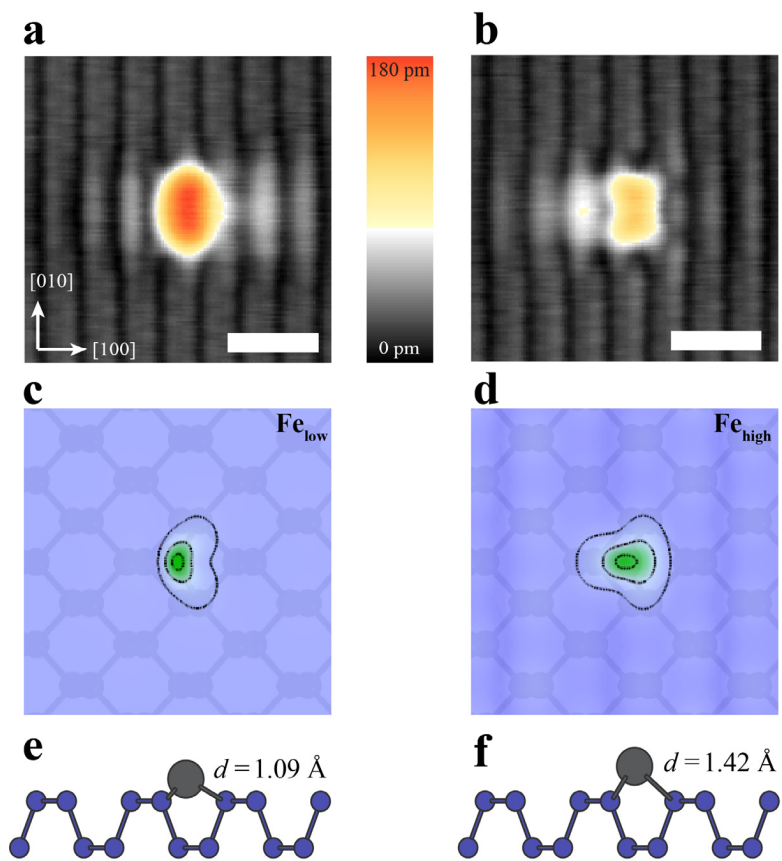

Figure 1. (a) STM constant-current image of Fe$_{low}$. (b) STM constant-current image of Fe$_{high}$. (c) DFT calculation of the expected charge density distribution for Fe$_{low}$, with the relaxed atomic structure (e). (d) DFT+U (U = 3eV) calculation of the expected charge density distribution for Fe$_{low}$, and the relaxed atomic structure (f).

| Valency | $n_d$ | $n_s$ | $m$ |
|---|---|---|---|
| Fe$_{low}$ | 7.54 | 0.24 | 2.0 µ$_B$ |
| Fe$_{high}$ | 7.07 | 0.46 | 3.5 µ$_B$ |

Table 1. Orbital occupation ($n_d$ and $n_s$) and magnetic moment ($m$) for each Fe valency.

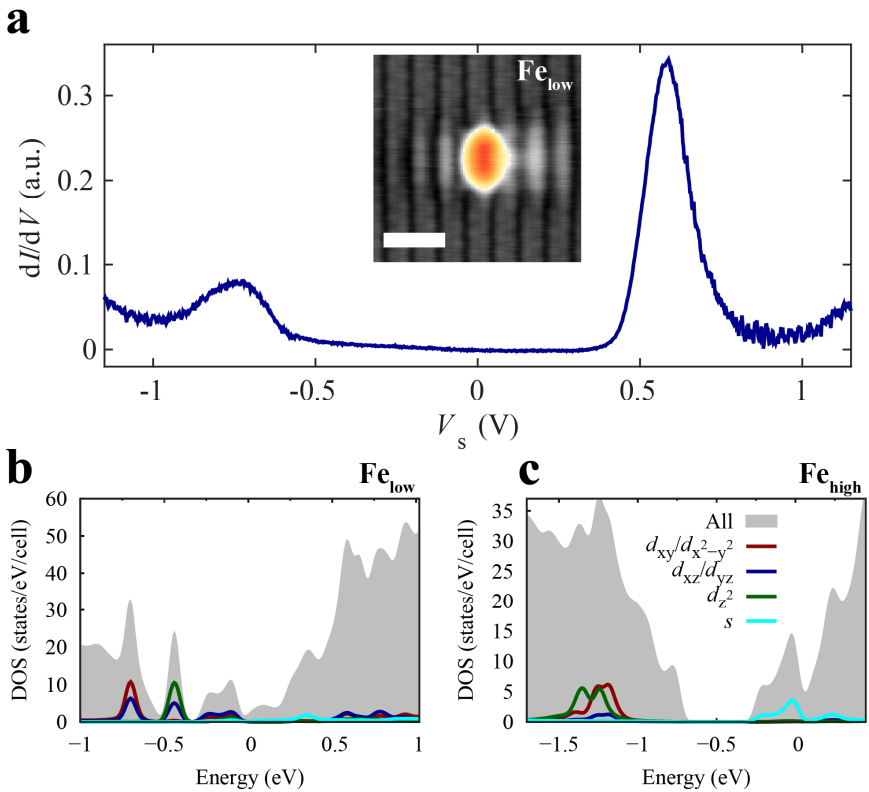

Figure 2. (a) Tunneling spectroscopy reveals the presence of two peaks in the STS on the Fe adatom. The peak energies are approximately -700 mV and 580 mV. (b) DFT+U calculation for the low-spin Fe state with U = 1.5 eV, revealing Fe based peaks in the DOS at energies of -700 mV, -500 mV, and -300 mV. (c) DFT+U calculation for the high-spin Fe state with U = 3 eV.

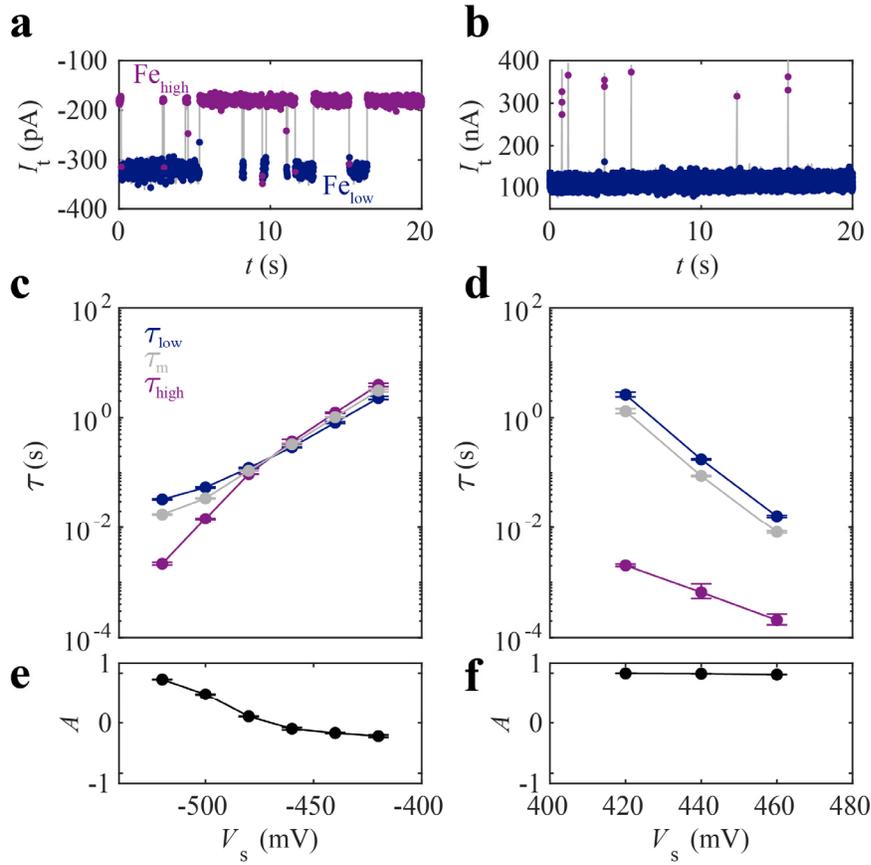

Figure 3. (a), (b) Current traces of the two-state telegraph signal with the atom switching stochastically between the two valencies at (a) $V_s = -420$ mV and (b) $V_s = 420$ mV. (c), (d) Residence lifetimes of $Fe_{low}$ (blue) and $Fe_{high}$ (purple) states, in addition to the mean residence lifetime (gray), $\tau_m = (\tau_{low} + \tau_{high})/2$, as a function of applied bias. (e), (f) Asymmetry, $A = (\tau_{low} - \tau_{high})/(\tau_{low} + \tau_{high})$ of the valencies with varying bias.

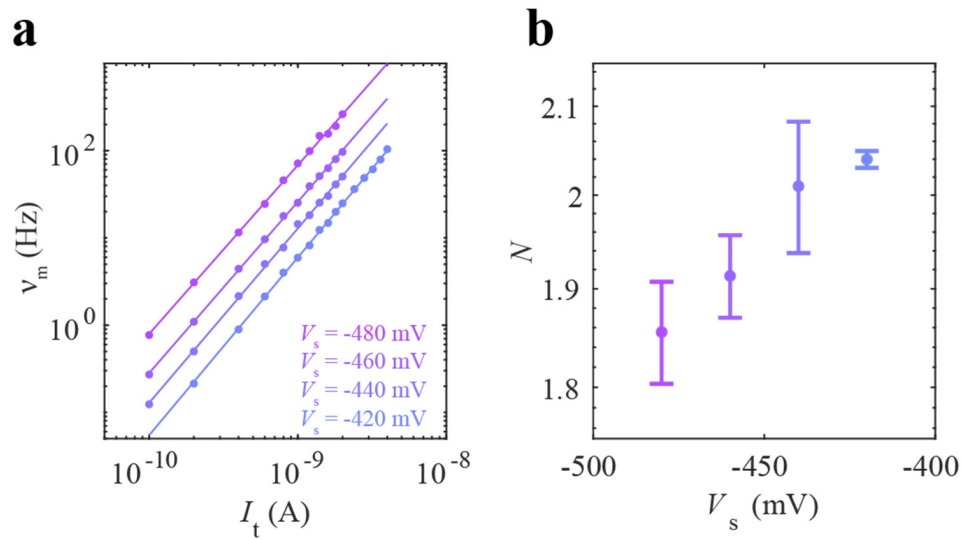

Figure 4. (a) Mean switching frequency as a function of current (both plotted using a logarithmic scale) for four different applied biases. (b) Power-law exponents ($\nu_m = I^N$) to the various data sets. The data cluster around $N=2$, indicative of a two-electron process involved in overcoming the ground state barrier.

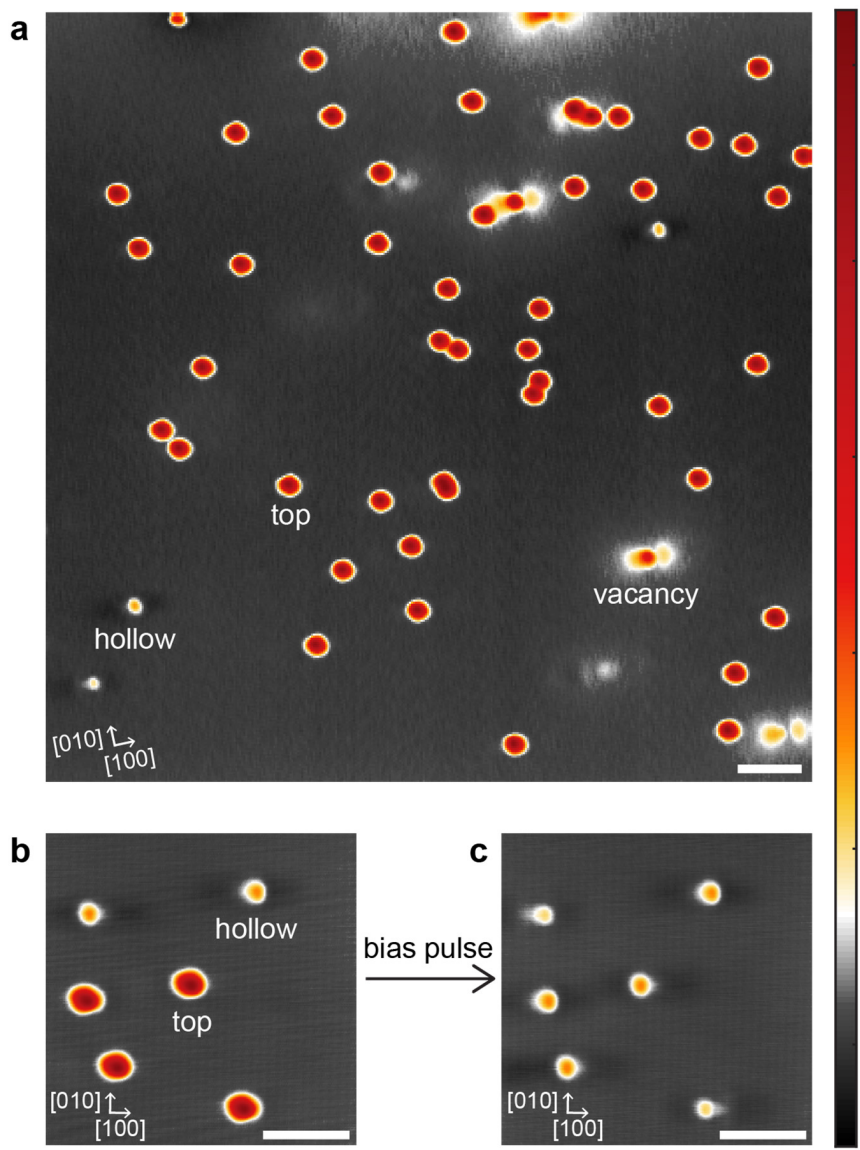

Figure 5. (a) Representative overview image of a BP sample after deposition of Fe atoms. (b)-(c) Fe atoms adsorbed on a top site can irreversibly be pushed into a hollow site by a bias pulse. In (c), both hollow site species $Fe_{low}$ and $Fe_{high}$ are visible. Parameters for (a)-(c): $V_s$ = -400 mV, $I_t$ = 20-60 pA, scale bar = 5 nm, $\Delta z$ = 290 pm.

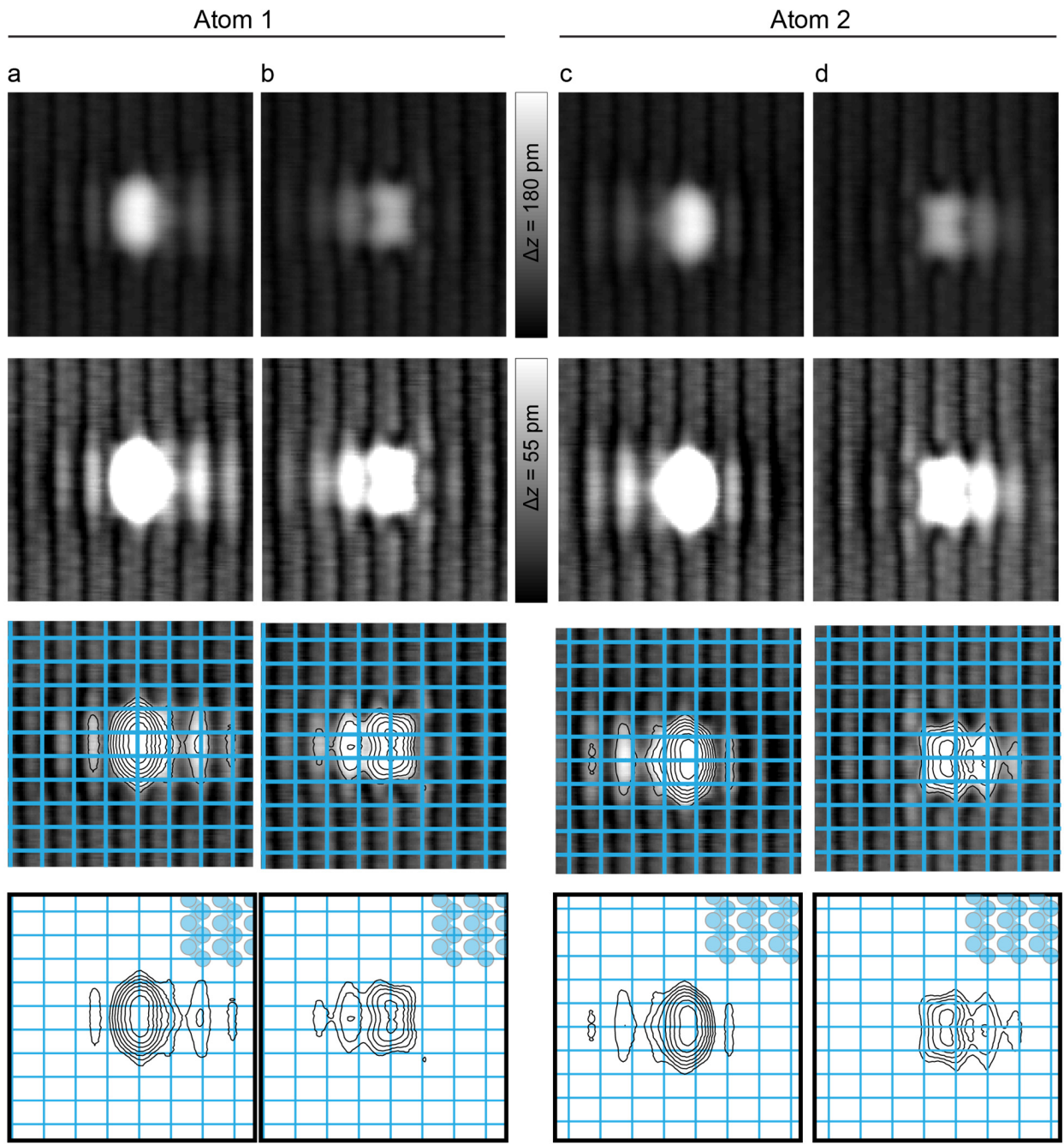

Figure 6. Binding site analysis for two different Fe atoms. (a) Binding site analysis for atom 1 in the low-spin state and (b) same analysis in the high-spin state. (c) and (d) show the binding site analysis for a second Fe atom residing in a mirror symmetric hollow site in the low-spin and high-spin valencies, respectively. Images taken with $V_s$ = -60 mV, $I_t$ = 100 pA.

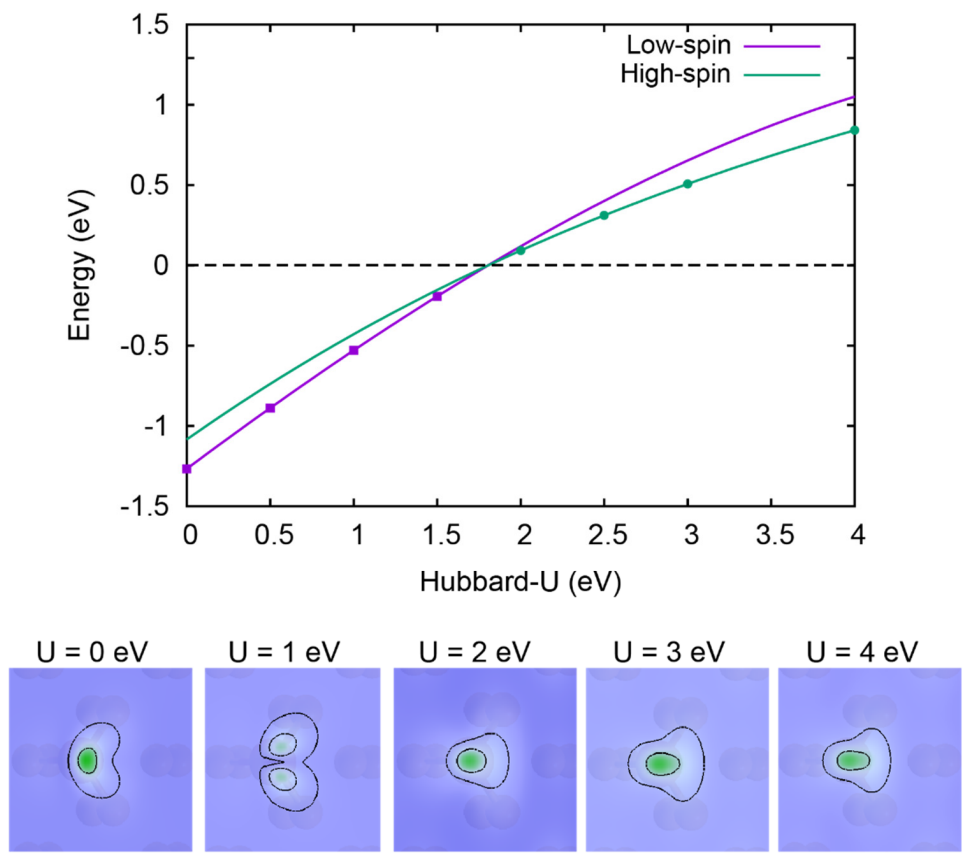

Figure 7. Calculations with varying Hubbard-U correction. The upper plot shows the total energy (markers) of the $Fe_{low}$ state (purple) for U < 1.8 eV and the $Fe_{high}$ state (green) for U > 1.8 eV. Solid lines show fits to the markers. The critical value for the U parameter is U = 1.8 eV, marking a point where both states are stable and nearly degenerate in energy. The lower panels show the calculated charge density distributions for varying Hubbard-U parameters.

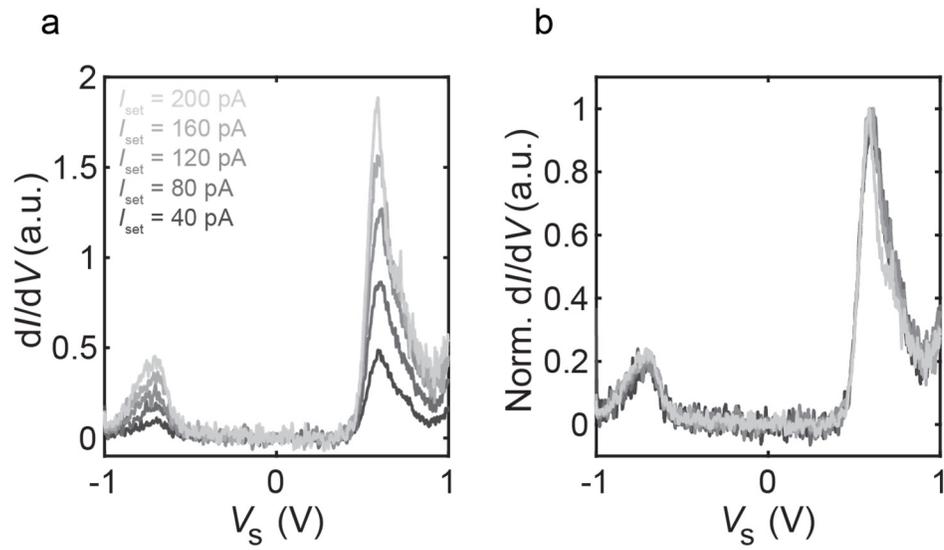

Figure 8. Setpoint dependent d*I*/d*V* spectra on the Fe$_{low}$ atom, with the raw spectra in (a) and normalized spectra in (b). Grayscale color used to indicate the varying current setpoint ($I_{set}$) used to modify the tip-sample separation.

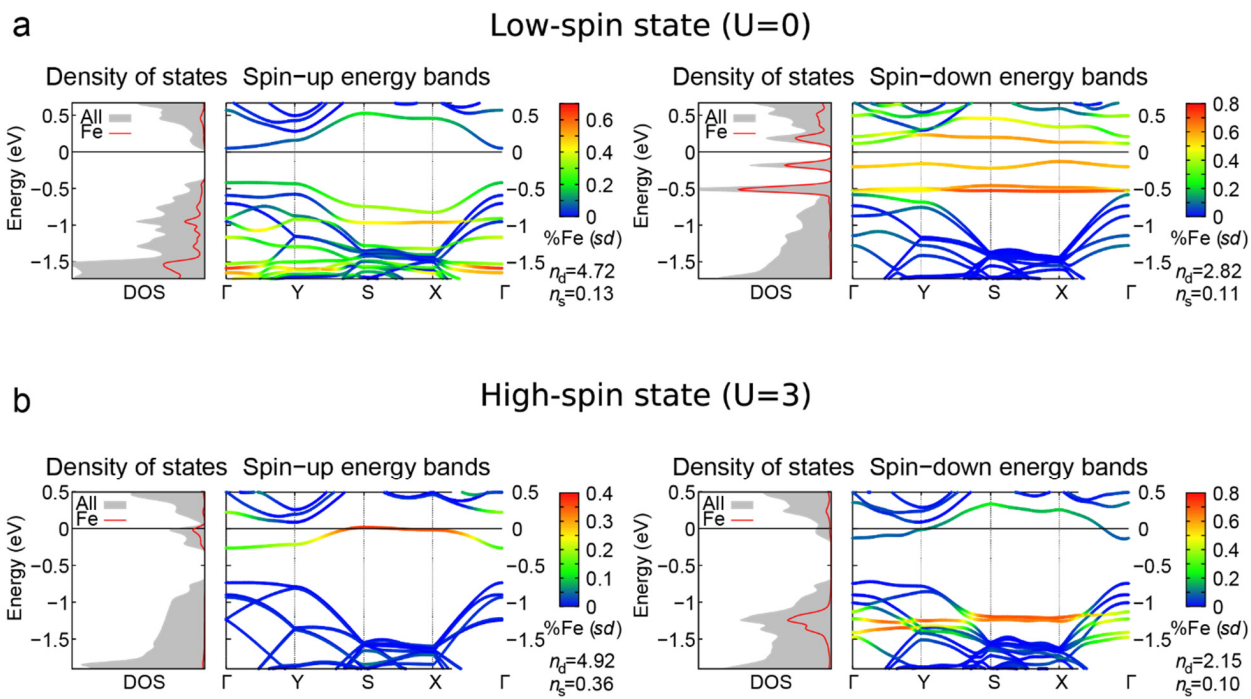

Figure 9. Band structure and density of state calculations for both Fe valencies. Spin resolved DOS and band structure for the low-spin valency (a) and the high-spin valency (b). The color scale marks the contribution of the different orbitals: blue bands stem exclusively from P orbitals, while red indicates a contribution from Fe orbitals. The occupation of the $3d$ and $4s$ shell of Fe are given below the color bar.

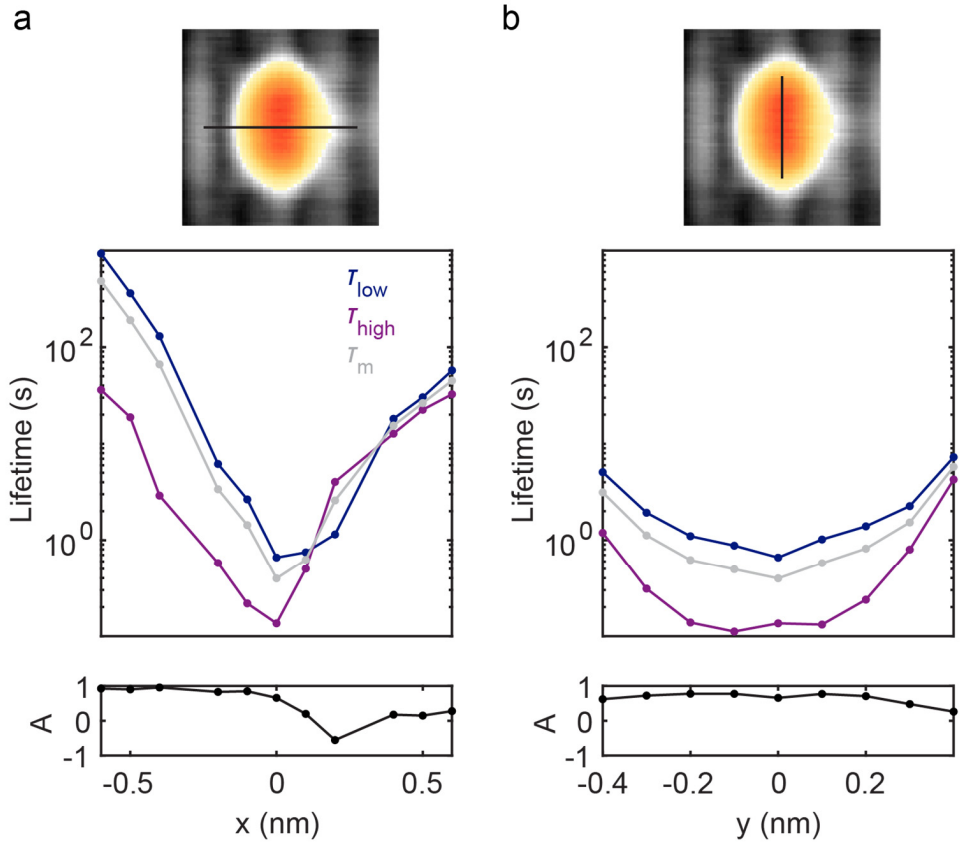

Figure 10. State lifetimes $\tau_{low}$ and $\tau_{high}$, mean lifetime $\tau_m = (\tau_{low}+\tau_{high})/2$ and asymmetry $A = (\tau_{low}-\tau_{high})/(\tau_{low}+\tau_{high})$ as a function of position, in the (a) [100] direction and (b) [010] direction (as marked in the STM images in the top panels). Measurements taken at $V_s = -460$ mV at constant height.

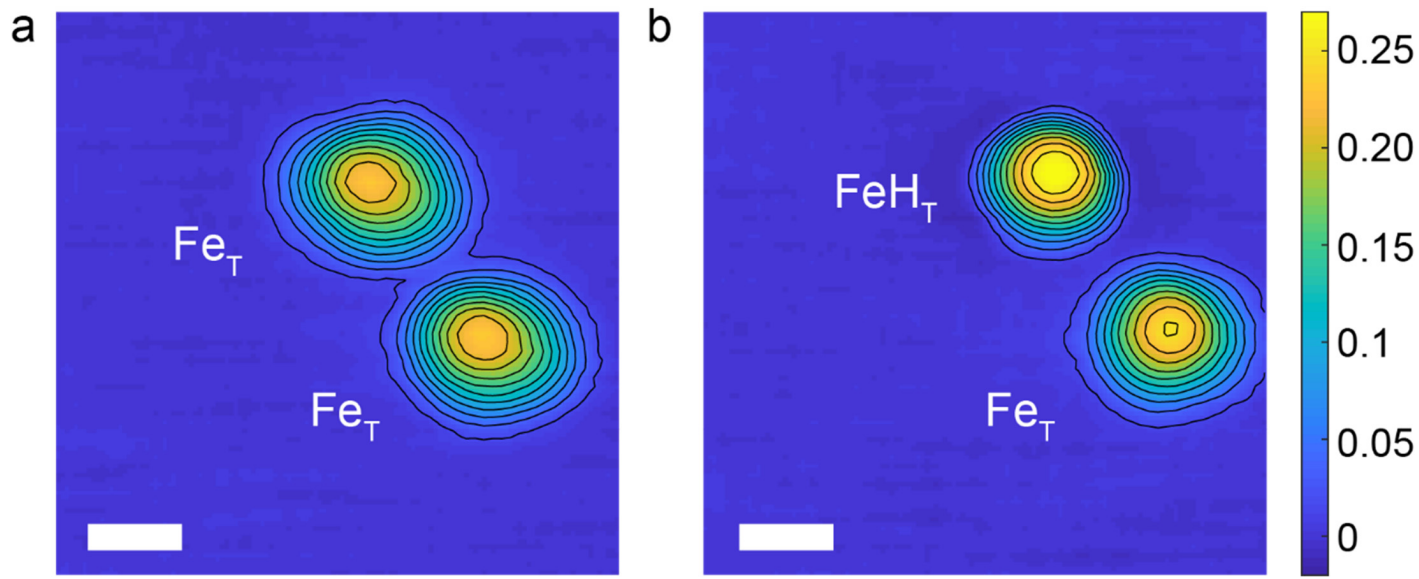

Figure 11. (a) Contour plot of an STM constant-current image of two Fe atoms in the top binding site (Fe$_T$) with an apparent height of 250 pm ($V_s$ = -400 mV, $I_t$ = 20 pA, scale bar = 1 nm). (b) Contour plot of STM constant-current image of the same two atoms, where one Fe atom is hydrogenated (FeH$_T$), distinguished by its larger apparent height of 270 pm ($V_s$ = -400 mV, $I_t$ = 20 pA, scale bar = 1 nm).